\documentclass[11pt,aps,prl,amsmath,amssymb,amsfonts,nofootinbib,long,floatfix]{revtex4}
\usepackage{graphicx,epsfig,latexsym,bm}

\newcommand\kk{\vec{k}}
\newcommand\qq{\vec{q}}
\newcommand\ka{\vec{\kappa}}

\begin{document}

\title{One-dimensional model of freely decaying two-dimensional turbulence}

\author{Leonardo Campanelli$^{1}$ \\
\footnotesize{email: \texttt{leonardo.s.campanelli@gmail.com}}}

%\email{leonardo.s.campanelli@gmail.com}

\affiliation{$^1$All Saints University, Asudom Academy of Science, 5145 Steeles Ave., Toronto (ON), Canada}

\date{\today}

%******************************   Abstract   *************************************************%

\begin{abstract}
\vspace{0.5cm}
\begin{center}
{\bf ABSTRACT}
\end{center}

We construct a discrete shell-model for two-dimensional turbulence that takes into
account local and nonlocal interactions between velocity modes in Fourier space.
In real space, its continuous limit is described by the one-dimensional
Burgers equation. We find a novel approximate scaling solution of such an equation
and show that it well describes the main characteristics of the energy spectrum in
fully developed, freely decaying two-dimensional turbulence.

\vspace{0.5cm}
Keywords: Two-dimensional turbulence; shell models; analytical methods

\end{abstract}

%*********************************************************************************************%

\maketitle

\newpage

%*********************************************************************************************%

\section{I. Introduction}

Freely decaying, two-dimensional hydrodynamic (2HD) turbulence has been the
object of intensive studies in the last few decades
(for a review of two-dimensional turbulence see, e.g.,~\cite{Boffetta}).
This is because, some three-dimensional turbulent systems in nature,
such as large-scale motions in the atmosphere and oceans,
are well approximated by two-dimensional hydrodynamical models.

The main features of statistically homogeneous and isotropic
turbulence are encoded in the so-called kinetic energy spectrum $E(k,t)$,
which defines the energy contained in a given (Fourier) velocity mode
of wavenumber $k$ at the time $t$. Such a spectrum exhibits,
in both direct numerical simulations and laboratory experiments,
peculiar ``universal'' properties that completely characterize the
evolution of fully developed turbulence. In particular,
the energy spectrum unveils the presence of three distinct
regions, or ranges, of turbulence with different characteristics:
the large-scale range, the inertial range, and the dissipative range.

The aim of this paper is to introduce and discuss a
reduced dimensional model of Navier-Stokes equation
that governs two-dimensional turbulence. Such a
model is, under plausible assumptions, analytically
solvable and seems to describe well the main properties
of freely decaying two-dimensional turbulence, namely
when external sources are not present in the turbulent state.

The plan of the paper is as follows.
In the following section, we review the main features of the energy spectrum
in freely decaying 2HD turbulence. In Section III, we apply
particular types of scaling arguments, first introduced by Olesen
in the study of freely decaying, three-dimensional magnetohydrodynamic turbulence, to
the hydrodynamical case in two-dimensions. In Section IV, we introduce
our one-dimensional model for two-dimensional turbulence.
In Section V, we find and discuss the solution of the equation
describing such a reduced dimensional model.
In Section VI, we draw our conclusion. Finally, in the Appendix, we
derive the Saffman spectrum from the Burgers equation.

\section{II. Hydrodynamics in two dimensions}

The evolution of an incompressible fluid in two dimensions is
described by the two-dimensional Navier-Stokes equation,
\begin{equation}
\label{NS}
\frac{\partial \textbf{v}}{\partial t} + (\textbf{v} \cdot \nabla)\textbf{v}
= -\nabla p + \nu \nabla^2 \textbf{v}
\end{equation}
and the incompressibility condition, $\nabla \cdot \textbf{v} = 0$,
where $\textbf{v}(\textbf{x},t) = (v_1,v_2)$ is
the velocity of bulk fluid motion, $p(\textbf{x},t)$ is the pressure,
and $\nu$ is the kinematic viscosity (see, e.g.,~\cite{Biskamp,DavidsonBook}).
One can define the kinematic Reynolds number, $\text{Re} = vl/\nu$,
where $v$ and $l$ are the typical velocity and typical
length scale of the fluid motion. Hydrodynamic turbulence occurs
when $\text{Re} \gg 1$.

In Fourier space,
\footnote{For the Fourier transform, we use the convention
$\textbf{v}(\textbf{x}) = \int \! \frac{d^n k}{(2\pi)^{n/2}} \, e^{i \textbf{k}\textbf{x}} \, \textbf{u}(\textbf{k})$
and
$\textbf{u}(\textbf{k}) = \int \! \frac{d^n x}{(2\pi)^{n/2}} \, e^{-i \textbf{k}\textbf{x}} \, \textbf{v}(\textbf{x})$.}
the two-dimensional hydrodynamic equations take the form
\begin{equation}
\label{dNS}
\left( \frac{\partial}{\partial t} + \nu k^2 \! \right) \!
u_{\alpha}(\textbf{k}) = i P_{\alpha \beta
\gamma}(\textbf{k}) \int \! d^{\,2} p \, u_{\beta}(\textbf{p}) \,
u_{\gamma}(\textbf{k} - \textbf{p})
\end{equation}
and $k_{\alpha} u_{\alpha} (\textbf{k},t) = 0$,
where $\textbf{u}(\textbf{k},t)$ is the velocity field in Fourier
space,
\begin{equation}
\label{P}
P_{\alpha \beta \gamma}(\textbf{k}) = \frac{1}{2\pi} \left( \frac{k_{\alpha} k_{\beta}
k_{\gamma}}{k^2} - \mbox{$\frac{1}{2}$} k_{\gamma} \delta_{\alpha
\beta} - \mbox{$\frac{1}{2}$} k_{\beta} \delta_{\alpha \gamma} \right) \!,
\end{equation}
and $k=|\textbf{k}|$. The pressure $p$ has been eliminated by use
of the incompressibility condition. Reality of the velocity field
in real space imposes that $\textbf{u}(-\textbf{k})=\textbf{u}^*(\textbf{k})$.
(Greek subscripts range from 1 to 2 and summation
over repeated indices is implied.)

For isotropic fluids, two-dimensional hydrodynamics
admits two invariants of motion in the limit of null kinematic
viscosity (the inviscid case): the kinetic energy density
(in a unitary two-dimensional volume),
\begin{equation}
\label{En}
E(t) = \langle \frac{1}{2} \int \! d^{\,2} x \,\textbf{v}^2 \rangle  =
\int_{0}^{\infty} \!\! dk E(k,t),
\end{equation}
and the enstrophy,
\begin{equation}
\label{enstrophy}
\Omega(t) = \langle \int \! d^{\,2} x \, \omega^2 \rangle
= \int_{0}^{\infty} \!\! dk k^2 E(k,t),
\end{equation}
where
\begin{equation}
\label{spectrum}
E(k,t) = \pi k \, \langle |\textbf{u}(\textbf{k})|^2 \rangle
\end{equation}
is the kinetic energy density spectrum, and
$\omega = \epsilon_{\alpha \beta} \partial_\alpha v_\beta =
\partial_1 v_2 - \partial_2 v_1$ is the ``scalar vorticity'' field.
Since we are interested in the evolution of statistically homogeneous
and isotropic velocity fields, an ensemble average denoted by
$\langle ... \rangle$ has been introduced in the above definitions
of energy and enstrophy.

The main features of locally homogeneous and isotropic
turbulence are described by the kinetic energy spectrum
to which we now turn our attention.
In freely decaying turbulence, the energy spectrum displays some
universal characteristics at both large and small scales. In
particular, the results of direct numerical simulations and
laboratory experiments clearly show the presence of three distinct
regions or ranges: ($i$) the large-eddies range,
$k \ll k_i$, ($ii$) the enstrophy inertial range,
$k_i \ll k \ll k_{diss}$, and the dissipation range, $k \gg k_{diss}$.
A sketch of the energy spectrum in these ranges is shown in Fig.~$1$.
The existence of such three regions is not surprising. Indeed,
in freely decaying 2HD there are only two independent length scales:
the dissipation length $L_{diss}(t) = \sqrt{\nu t}$ and
the initial scale $L_i$, which we assume to be much greater than
$L_{diss}$, $L_i \gg L_{diss}$. The wavenumber corresponding to $L_i$ is
$k_i = 1/L_i$, while the wavenumber corresponding to $L_{diss}$ is
$k_{diss} = 1/L_{diss}$. These two length scales define the three ranges
introduced above.

%*************************************   Figure 1  *******************************************%

\begin{figure}[t!]
\begin{center}
\includegraphics[clip,width=0.65\textwidth]{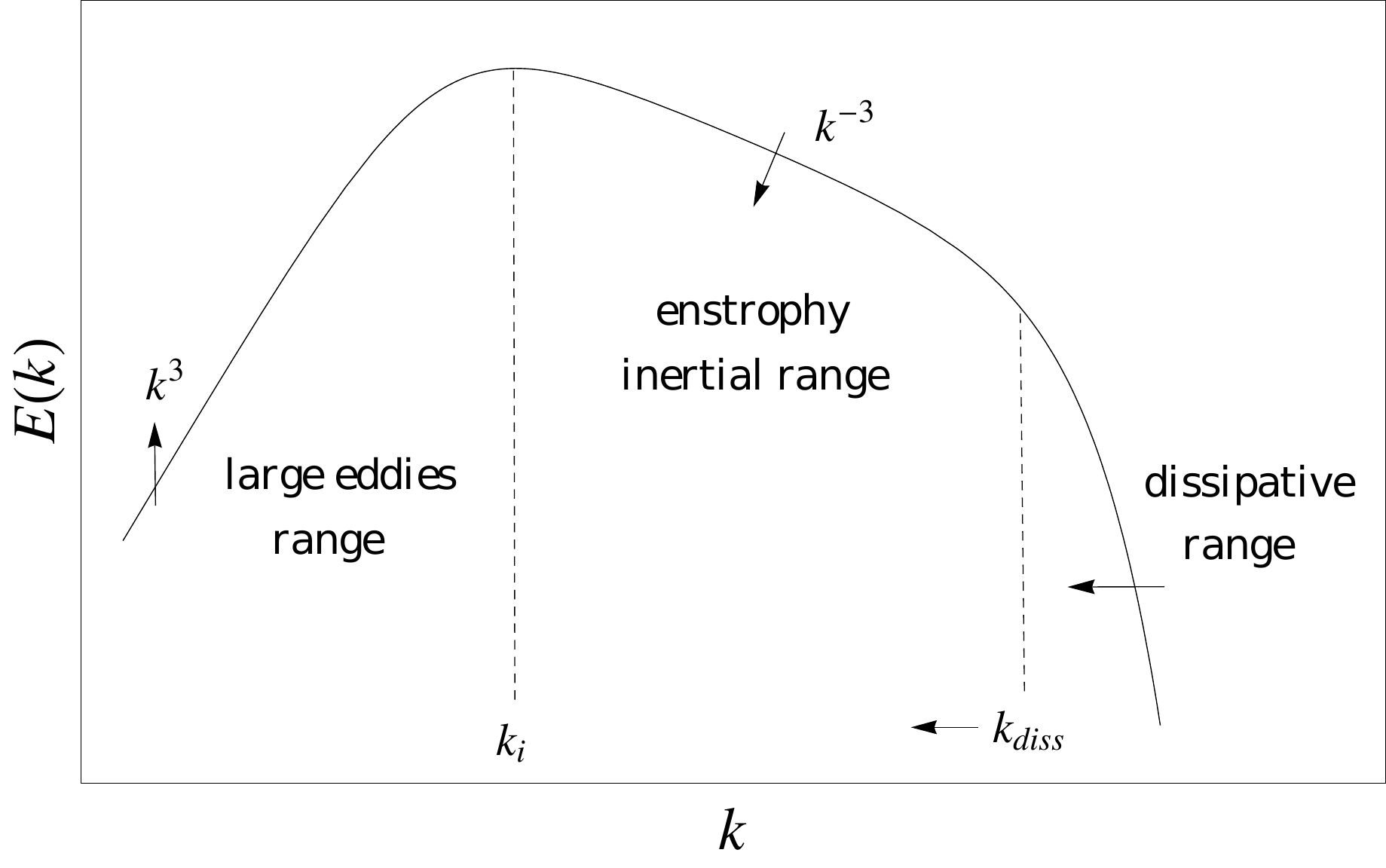}
\caption{Picture (log-log scale) of the expected energy spectrum $E(k,t)$ in freely
decaying two-dimensional turbulence as a function of the wavenumber $k$.
A time-dependent $k^3$ power-law spectrum
is shown at large scales ($k \rightarrow 0$). Arrows indicate displacements in time.}
\end{center}
\end{figure}

%*********************************************************************************************%

{\bf Energy inertial range.} -- A physical homogeneous and isotropic field not
interacting with external systems is dependent only upon the initial conditions which
introduces a mean velocity $u_i$ and a length scale $L_i$. This is the case, for example,
of turbulence behind a grid (with $u_i$ being the velocity field at the grid
and $L_i$ being the spacing of the bars of the grid perpendicular to the direction of the flow).
In numerical simulations, instead, it is assumed that the initial velocity field is
homogeneous and isotropic. In this case, the mean value of the velocity is zero and $u_i$ in the
following equations must be regarded as a (root-mean-square) r.m.s value of the field. More generally,
$L_i$ and $u_i$ can be considered as typical scale values at the onset of fully developed
turbulence.

By definition, an inertial range is a range where
dissipation is negligible and the dynamics is independent on the initial conditions.
Dimensional analysis, then, completely gives the shape of the
energy spectrum in such a range (see, e.g.,~\cite{Campanelli}),
\begin{equation}
\label{11}
E(k,t) = c t^{-2} k^{-3}, ~~~ k_i \ll k \ll k_{diss},
\end{equation}
where $c$ is a dimensionless constant.
In forced turbulence, a time-independent spectrum of the form $E(k,t) \propto k^{-3}$
is known as Batchelor spectrum~\cite{Batchelor}.

{\bf Dissipative range.} -- At small scales, $k \gg k_{diss}$, the 2HD equations
can be exactly solved if one assumes that the rate of energy transfer due to
nonlinear interactions is completely negligible.
Indeed, it can be shown that the energy spectrum changes in time as
$\partial E(k,t)/ \partial t = T(k,t) - \nu k^2 E(k,t)$,
where $T(k,t)$ is the energy change caused by nonlinear interactions
(see, e.g., ~\cite{Boffetta}).
When this term is negligible with respect to the viscous term, one obtains
\begin{equation}
\label{2}
E(k,t) = c_{\infty} \nu^2 k e^{-\nu k^2 t}, ~~~ k \gg k_{diss},
\end{equation}
where $c_{\infty}$ is a dimensionless constant.

The hypothesis that the rate of nonlinear energy transfer is
negligible is certainly true in the ``mathematical'' limit $k \rightarrow \infty$.
However, such a hypothesis is not justified in a region around the dissipation wavenumber.
Indeed, Tatsumi and Yanase~\cite{Tatsumi}, using the so-called ``modified zero
fourth-order cumulant approximation'',
found an analytic expression of the energy spectrum in the dissipation
range of the form
\begin{equation}
\label{6a}
E(k,t) \propto e^{- b k \sqrt{\nu t}}, ~~~ k \gtrsim k_{diss},
\end{equation}
where $b$ is a dimensionless constant. The same authors found, by
direct numerical integration of 2HD equations,
that $E(k,t) \propto \exp{\!\left[-b(k \sqrt{\nu t})^s \right]}$
with $s$ in the range
$[1.3,1.4]$. The discrepancy with Eq.~(\ref{6a}) indicates, according to the authors,
that either the asymptotic behaviour $s = 1$ is realized beyond the numerical coverage or
the numerical results are not accurate
enough. In either case, the asymptotic form of the spectrum is
different from the purely viscous spectrum~(\ref{2}) for
$k \gtrsim k_{diss}$.

{\bf Large-eddies range.} -- In the limit $k \rightarrow 0$,
a Maclaurin expansion of the energy spectrum gives (see, e.g.~\cite{DavidsonBook})
\begin{equation}
\label{expansion}
E(k,t) =
\left\{ \begin{array}{lll}
     4\pi \mathcal{L} k + \mathcal{O}(k^3) ,   & ~ \mathcal{L} \neq 0,   \\
     I(t) k^3 + \mathcal{O}(k^5),              & ~ \mathcal{L} = 0.
    \end{array}
\right.
\end{equation}
Here, $\mathcal{L}$ is known as the two-dimensional Saffman integral,
and its time-independence is a consequence of the conservation of
linear momentum~\cite{DavidsonBook}.
Indeed, Davidson~\cite{Davidson} has shown that the Saffman integral
in two dimensions can be written as
\begin{equation}
\label{L}
\mathcal{L} = \langle \frac{1}{V} \!
\left( \, \int_{V} \! d^{\,2} x \, \textbf{v} \right)^{\!2} \rangle,
\end{equation}
where $V$ is some large two-dimensional volume. Accordingly,
$\mathcal{L}$ is nonzero whenever the turbulence has sufficient
linear momentum. In this case, then, the energy spectrum at large scales is
linear in the wavenumber, a result known
as Saffman spectrum. If the total linear momentum is instead negligible,
a spectrum of the form $I(t) k^3$ develops at large scales where
$I(t)$, the so-called Loitsyansky integral in two dimensions, is
generally expected to be an increasing function of time~\cite{DavidsonBook}.

\section{III. Dimensional scaling}

{\bf Olesen's scaling.} -- The two-dimensional HD equations, under the scaling
transformations
%
%$\textbf{x} \rightarrow \ell \, \textbf{x}$, $t \rightarrow \ell^{1-h} \, t$,
%
\begin{equation}
\label{Ole0}
\textbf{x} \rightarrow \ell \, \textbf{x}, ~~ t \rightarrow \ell^{1-h} \, t,
\end{equation}
admit solutions of the type
\begin{equation}
\label{Ole1}
\textbf{v}(\ell \, \textbf{x},\ell^{1-h} \, t) = \ell^{\,h} \, \textbf{v}(\textbf{x},t),
\end{equation}
provided that the dissipative parameter $\nu$ scales as
\begin{equation}
\label{Ole2}
\nu(\ell^{1-h} \, t) = \ell^{1+h} \, \nu(t).
\end{equation}
Here $\ell > 0$ is  the ``scaling factor'' and $h$ is a arbitrary
real parameter. Starting from the scaling relations,
and following Olesen's analysis~\cite{Olesen}, we obtain the energy spectrum
\begin{equation}
\label{Ole3}
E(k,t) = k t^{2p} \Psi(k t^{(1+p)/2}),
\end{equation}
where $p = (1+h)/(1-h)$ and $ \Psi$ is a arbitrary scaling-invariant function
of its argument. We have two cases:

($i$) If dissipation is negligible, as it happens at large scales and,
by definition, in the inertial range, then $h$ (and in turns $p$) is
completely arbitrary, to wit, it is not fixed by scaling arguments.

($ii$) If dissipation is important, as in the dissipative range (at very small scales),
then $h$ is fixed by the scaling properties of $\nu$.
Indeed, if we differentiate Eq.~(\ref{Ole2}) with respect to
$\ell$, and put $\ell = 1$ afterwards, we get $\nu \propto t^p$.
In particular, as we will assume in this paper, if $\nu$ is constant
then we must take to $p = 0$ and, accordingly,
\begin{equation}
\label{Ole4}
E(k,t) = k \psi(k t^{1/2}).
\end{equation}
If dissipation is not negligible and $\nu$ does not evolve in time as a power law,
then Eq.~(\ref{Ole1}) is not a solution of the 2HD equations and in turn the
energy spectrum do not exhibit a scaling of the type given by Eq.~(\ref{Ole3}).

{\bf Dimensional analysis.} -- For the sake of later convenience let us
re-write Eqs.~(\ref{Ole4}) and (\ref{Ole3}) in a form such that the
scaling functions $\Psi$ and $\psi$ are dimensionless.
It is easy to check that they must be written as
\begin{equation}
\label{Ole5}
E(k,t) = \nu^2 k \psi(k\sqrt{\nu t})
\end{equation}
and
\begin{equation}
\label{Ole6}
E(k,t) = E_i \kappa \tau^{2p} \Psi(\kappa \tau^{(1+p)/2}),
\end{equation}
respectively. Here, $E_i = u_i^2 L_i$, $\kappa = k/k_i$, $\tau = t/t_i$,
and $t_i = L_i/u_i$ is the so-called initial ``eddy turnover time''.

It is interesting to observe that form of the energy spectrum in the inertial range
is completely fixed by scaling considerations only. Indeed, since in this range
the dynamics is completely independent on dissipation (and on the initial conditions),
the only possible form for the scaling function $\psi$ such that the energy spectrum
does not depend on the kinematic viscosity is $\psi(x) = c x^{-4}$. This gives
the Batchelor spectrum in Eq.~(\ref{11}) with $c$ being a ``universal''
dimensionless constant.

\section{IV. One-dimensional model}

In this Section, we propose a one-dimensional, continuous (toy) model for
two-dimensional hydrodynamics. We start by generalizing the well-known
discrete shell model~\cite{shell} for HD turbulence by including
all possible local and nonlocal interaction terms
in Fourier space. Next, we consider the continuous limit of such a discrete shell model.
The resulting model will be analyzed in the next section.

{\bf Shell model with nonlocal interactions.} -- Let us consider a
shell model constructed in a discrete wavenumber space which is
approximated by $2N$ shells,
%$\{ k_n \, | \, k_n = \! Kn, \; n = \! -N,-N+1,...,0,...,N-1,N \}$.
%
\begin{equation}
\label{shell0}
\{ k_n \, | \, k_n = \! Kn, \; n = \! -N,-N+1,...,0,...,N-1,N \},
\end{equation}
where $K$ gives the spacing between two consecutive shells.
The velocity is represented by a set
of complex variables $\{ u_n \}$, where $u_n$ stands for the
velocity components whose scalar wavenumber $\kk \in \mathbb{R}$ satisfies $k_n \leq \kk
\leq k_{n+1}$. We now construct the
Navier-Stokes equation in this scalar model in the following way,
\begin{equation}
\label{shell1}
\left( \frac{d}{d t} + \nu k_n^2 \right) \! u_n = i P_n \!
\sum_{j=-N}^{N} \! u_j \, u_{n-j},
\end{equation}
where $P_{n}$ is a real vector and $u_{-n} = u_n^*$.
%
%\footnote{The phase space is a real $4N$-dimensional space $\{ u_n^R,u_n^I\}$,
%where $u_n^R$ and $u_n^I$ are the real and imaginary parts of $u_n$, respectively.
%It is straightforward to check that the phase
%volume conservation holds
%in the inviscid case, namely $\sum_{n}(\partial \dot{u}_n^R /
%\partial u_n^R + \partial \dot{u}_n^I / \partial u_n^I) = 0$ for $\nu = 0$,
%where a dot denotes a time derivative.}
%
The sum on the left-hand side of~(\ref{shell1}) includes all possible
local and nonlocal interaction terms between different Fourier modes.

In the shell model, the kinetic energy density is
$E(t) = \sum_{n=0}^N  E_{|n|}$, where $E_{|n|}(t) = \pi k_{|n|} |u_n|^2$
is the energy spectrum (for a justification of the form of the energy spectrum,
see below.)
One can check that a sufficient condition for the conservation of energy in
the inviscid limit is $P_{-n} = -P_{n}$.
Mimiking Eq.~(\ref{P}), we can assume assume that $P_{n}$ is linear in $n$,
$P_{n} \propto k_{n}$. With this $P_{n}$, it is easy to check that the quantity
$\sum_{n=0}^N  f(|k_n|) \, |u_n|^2$, with $f(x)$ any function such that
$f(0) = 0$, is conserved in the limit of vanishing viscosity.
In particular, for $f(x) = x^3$ the above expression is the analogue of the enstrophy.
%This suggests that the shell model above introduced
%is an appropriate model of two-dimensional hydrodynamic turbulence.

{\bf Continuous limit.} -- The continuous limit of the above shell model
corresponds to take $N \rightarrow \infty$ and $K \rightarrow 0$. In this case,
$k_n$ is replaced by $\kk$, the discrete set $u_n(t)$ is replaced by a continuous
complex variable $u(\kk,t)$, the real vector $P_{n}$ by a real function of $P(\kk)$, and
sums are replaced by integrals, e.g.,
$K \sum_{j=-N}^{+N} \rightarrow \int_{-\infty}^{+\infty} d\kk$.
%
%\begin{equation}
%\label{shell5}
%k_n \rightarrow \kk, ~~~ u_n(t) \rightarrow u(\kk,t), ~~~ P_{n} \rightarrow P(\kk),
%~~~ K \! \sum_{j=-N}^{+N} \rightarrow \int_{-\infty}^{+\infty} d\kk.
%\end{equation}
%
The continuous limit of the discrete shell model~(\ref{shell1}) is then given by
\begin{equation}
\label{cNS}
\left( \frac{\partial }{\partial t} + \nu k^2 \right) \! u(\kk) = i P(\kk) \!
\int_{-\infty}^{+\infty} \!\! d\qq \, u(\qq) \, u(\kk-\qq),
\end{equation}
where $k = |\kk|$, $P(\kk) \propto \kk$, and $u(-\kk) = u^*(\kk)$. The energy spectrum is
\begin{equation}
\label{xyz}
E(k,t) = \pi k \, |u(\kk)|^2
\end{equation}
%
%(for a justification of the above definition, see below),
and the energy is $E(t) = \int_{0}^{\infty} \! dk E(k,t)$.
The energy, and in the general the quantity
$\int_{0}^{\infty} \! dk g(k) \, |u(\kk)|^2$ with $g(k)$ being any real function such that
the previous integral exists, is conserved in the limit of vanishing viscosity.

In the following, we take
\begin{equation}
\label{PPP}
P(\kk) = -\kk/2\sqrt{2\pi}.
\end{equation}
%
%$P(\kk) = -\kk/2\sqrt{2\pi}$.
A possible multiplicative constant in the above definition of $P(\kk)$
can be reabsorbed in the definition of time and kinematic viscosity,
and then will be neglected in the following.

{\bf Burgers equation.} -- The factor $-1/2\sqrt{2\pi}$ in Eq.~(\ref{PPP})
has been introduced for convenience. In this case, indeed, the inverse
Fourier transform of Eq.~(\ref{cNS}) is the well-known viscid Burgers equation
(see~\cite{Burgers} and references therein)
\begin{equation}
\label{Burgers}
\frac{\partial v}{\partial t} +
v \frac{\partial v}{\partial \zeta} = \nu \frac{\partial^2 v}{\partial \zeta^2},
\end{equation}
where $v(\zeta,t) = \int \! \frac{d \kk}{\sqrt{2\pi}} \, e^{i \kk \zeta} u(\kk,t)$.

It is assumed in the literature that the viscid Burgers equation cannot describe the main features
of hydrodynamic turbulence~\cite{Burgers}. This is because an exact solution of such an equation
can be given that does not share an important
property of the Navier–Stokes equation, namely the sensitivity to small changes in the initial conditions (at sufficiently high Reynolds numbers) that trigger
the spontaneous arise of randomness by chaotic dynamics.
Such an exact solution is found by applying a
Cole-Hopf transformation defined by introducing the function $f(\zeta,t)$ as
\begin{equation}
\label{Cole-Hopf}
v = - \frac{2\nu}{f} \, \frac{\partial f}{\partial \zeta} \, .
\end{equation}
Inserting the above equation in the viscid Burgers equation, one gets
\begin{equation}
\label{Burgers2}
H \, \frac{\partial f}{\partial \zeta} = f \frac{\partial H}{\partial \zeta} \, ,
\end{equation}
where
\begin{equation}
\label{heat}
H = \frac{\partial f}{\partial t} - \nu \frac{\partial^2 f}{\partial \zeta^2} \, .
\end{equation}
Consequently, if $f$ solves the ``heat equation'' $H = 0$
than $v$, given the Cole-Hopf transformation, solves the viscid Burgers equation.
Since the heat equation can be exactly solved, the exact form of $f$ is known.
An inverse Cole-Hopf transformation then gives the analytical
expression for $v$~\cite{Burgers}. Moreover, it can be shown that
such a solution for $v$ is unique if the initial condition
\begin{equation}
\label{initial}
v(\zeta,0) = v_0(\zeta)
\end{equation}
is given. The existence of such a solution, then, excludes the possibility that
the Burger equation could describe the {\it transition} from an initial state
to a state of fully developed hydrodynamic turbulence.

However, the above arguments do not exclude the possibility that
the Burger equation could describe the evolution of the system
{\it after} a turbulent state has been reached. Indeed,
our claim is that this is the case, in the sense that
Burgers equation does describe the main features of the energy spectrum in
fully developed, freely decaying two-dimensional turbulence. In order this to be possible
one has to assume that the Burgers equation is a model of turbulence
only at times $t \gg t_i$, namely only when turbulence is already fully developed.
In particular, the initial condition~(\ref{Burgers}) is inapplicable in this case,
which means that the velocity field must be not defined for $t=0$
(or, in general, for $t=t_0$, where $t_0$ is the initial time).

A possible solution of the Burgers equation that satisfies the above condition and that
will correspond, as discussed in Section V, to the case of large wavenumbers can
be obtained by writing $v(\zeta,t)$ as
\begin{equation}
\label{Burgers3}
v(\zeta,t) = \sqrt{\frac{\nu}{t}} \, \hat{\phi}\!\left( \! -\frac{\zeta}{\sqrt{\nu t}} \right) \!,
\end{equation}
with $\hat{\phi}(x)$ a ``scaling function'' to be determined.
Observe that this solution is not defined for $t=0$.

Inserting into Eq.~(\ref{Burgers}) we get
\begin{equation}
\label{Burgers4}
2\hat{\phi}'' + 2 \hat{\phi} \hat{\phi}' + \omega \hat{\phi}' + \hat{\phi} = 0,
\end{equation}
where a prime indicates a derivative with respect to $\omega = -\zeta/\sqrt{\nu t}$.
The solution of the above equation is
\begin{equation}
\label{Burgers5}
\hat{\phi}(\omega) = \frac{2c_1c_2 H_{c_1-1} \!\! \left(\frac{\omega}{2}\right)
- \omega \! \left[c_2 H_{c_1} \!\! \left(\frac{\omega}{2}\right) + \, _1F_1 \!\!\left(\!-\frac{c_1}{2},\frac12;\frac{\omega^2}{2} \right) + c_1 \, _1F_1 \!\! \left(1-\frac{c_1}{2},\frac32; \frac{\omega^2}{2} \right)\right]}{c_2 H_{c_1-1} \!\! \left(\frac{\omega}{2}\right) + \, _1F_1 \!\!\left(\!-\frac{c_1}{2},\frac12;\frac{\omega^2}{2} \right)},
\end{equation}
%
%\begin{equation}
%\label{Burgers5}
%\hat{\phi}(\omega) = \frac{2c_1c_2 h_1 - \omega (c_2 h_2 + f_1 + c_1 f_2)}{c_2 h_1 + f_1} \, ,
%\end{equation}
%
where $c_1$ and $c_2$ are real constants of integration,
%
%\begin{eqnarray}
%&& h_1(\omega) = H_{c_1-1} \!\! \left(\frac{\omega}{2}\right) \!, \nonumber \\
%&& h_2(\omega) = H_{c_1} \!\! \left(\frac{\omega}{2}\right) \!, \nonumber \\
%&& f_1(\omega) = \,_1F_1 \!\!\left(\!-\frac{c_1}{2},\frac12;\frac{\omega^2}{2} \right)\!, \nonumber \\
%&& f_2(\omega) = \, _1F_1 \!\! \left(1-\frac{c_1}{2},\frac32; \frac{\omega^2}{2} \right) \!, \nonumber
%\end{eqnarray}
%
$H_n(x)$ is the Hermite polynomial of degree $n$, and $_1F_1(a,b;x)$ is the Kummer
confluent hypergeometric function~\cite{Abramowitz}.

As we will discuss later, the solution~(\ref{Burgers}) describes
the system in a state of fully developed HD turbulence.
The constants $c_1$ and $c_2$ are then to be considered ``universal''
and cannot be determined by initial conditions. Indeed, they will corresponds
to the constants $C$ and $c_\infty$ defined in Section V.

Let us finally observe that the function $f(\zeta,t)$ defined by
the Cole-Hopf transformation does not satisfy the heat equation.
Indeed, writing
\begin{equation}
\label{Burgers6}
f(\zeta,t) = \sqrt{\frac{\nu}{t}} \, \hat{\varphi}\!\left( \! -\frac{\zeta}{\sqrt{\nu t}} \right) \!,
\end{equation}
where $\hat{\varphi}(x)$ is an arbitrary function of its argument, and inserting into the Cole-Hopf transformation~(\ref{Cole-Hopf}), we get
$\hat{\phi} = 2 \hat{\varphi}'/\hat{\varphi}$, from which
\begin{equation}
\label{Burgers7}
\hat{\varphi} (\omega) = e^{\frac12 \! \int \! d\omega \hat{\phi}},
\end{equation}
with $\hat{\phi}$ given by Eq.~(\ref{Burgers5}). It is easy to check now,
by direct inspection, that $f(\zeta,t)$ as given by Eqs.~(\ref{Burgers6}) and
(\ref{Burgers7}) does not satisfy the heat equation.

In the next paragraph we show that, under particular
conditions, the Burger equation describes the dynamics
of a two-dimensional fluid. Moreover, the results of this analysis
will provide a justification of the definition of energy spectrum
introduced above. In the next section, instead,
we will discuss in some detail the above
``scaling solution'' of Burger equation. For the sake of convenience,
we will work in Fourier space and then consider Eq.~(\ref{cNS}).

{\bf Compressible fluid with negligible pressure.} -- Let us
consider a compressible fluid with negligible pressure.
We introduce the ``complex velocity field'' $V(x,y)$ as
%$V = v_x + i v_y$.
%
\begin{equation}
\label{x1}
V(x,y) = v_x + i v_y.
\end{equation}
Using the Navier-Stokes equation, we find that
the $V$ satisfies the equation
\begin{equation}
\label{x2}
\frac{\partial V}{\partial t} +
\left(\mbox{Re}[V] \frac{\partial }{\partial x}
+ \mbox{Im}[V] \frac{\partial }{\partial y}\right) \! V
= \nu \nabla^2 V.
\end{equation}
Taking the complex conjugate of the above equation, %we get
%
%\begin{equation}
%\label{x3}
%\frac{\partial V^*}{\partial t} +
%\left(\mbox{Re}[V] \frac{\partial }{\partial x}
%+ \mbox{Im}[V] \frac{\partial }{\partial y}\right) \! V^* = \nu \nabla^2 V^*.
%\end{equation}
%
it is easy to see that a possible solution of it is
such that $V^* = e^{-i \varphi} V$,
%
%\begin{equation}
%\label{x4}
%V^* = e^{-i \varphi} V
%\end{equation}
%
with $\varphi \in \mathbb{R}$. %This corresponds to $v_y = \tan(\varphi/2) \, v_x$.
%
%\begin{equation}
%\label{x5}
%v_y = \tan(\varphi/2) \, v_x
%\end{equation}
%

Let us introduce the new real quantity $\upsilon(x,y)$ as
\begin{equation}
\label{x6}
\upsilon(x,y) = e^{-i\varphi/2} V = \frac{1}{2} \left[ \sec(\varphi/2) v_x + \csc(\varphi/2) v_y \right].
\end{equation}
It satisfies the equation
\begin{equation}
\label{x7}
\frac{\partial \upsilon}{\partial t} +
\upsilon\left[\cos (\varphi/2) \frac{\partial }{\partial x} +
\sin (\varphi/2) \frac{\partial }{\partial y}\right] \!\upsilon
= \nu \nabla^2 \upsilon.
\end{equation}
We can diagonalize the operator in square parenthesis in the above equation
by performing a rotation of an angle $\varphi$ of the coordinate system,
\begin{eqnarray}
&& \xi = \cos (\varphi/2) x + \sin (\varphi/2) y, \\
&& \eta = -\sin (\varphi/2) x + \cos (\varphi/2) y.
\end{eqnarray}
The function $\upsilon$ in the new coordinate system, $\upsilon(\xi,\eta)$,
satisfies the equation
\begin{equation}
\label{x8}
\frac{\partial \upsilon}{\partial t} +
\upsilon \frac{\partial \upsilon}{\partial \xi} = \nu \nabla^2 \upsilon,
\end{equation}
where $\nabla^2 = \partial_\xi^2 + \partial_\eta^2$.
The r.m.s. value of the velocity field, $v_{rms}$, is given by
\begin{equation}
\label{x9}
v_{rms}^2(t) =  \int \! d^{\,2} x \, \textbf{v}^2(\textbf{x})
=\int \! d^{\,2} \boldsymbol{\zeta} \upsilon^2(\boldsymbol{\zeta})
= \int \! dk \, w_{rms}^2(k),
\end{equation}
where $\boldsymbol{\zeta} = (\xi,\eta)$ and
\begin{equation}
\label{x10}
w_{rms}^2(k,t) = 2\pi k w(\textbf{k}) \, w^*(\textbf{k}),
\end{equation}
with $w_{rms}$ being the spectrum of the r.m.s. velocity.
Here, $w(\textbf{k})$ is the
Fourier transform of $\upsilon(\boldsymbol{\zeta})$,
$\textbf{k} = (k_\xi,k_\eta)$, and $k = |\textbf{k}|$.

Let us now assume that $\upsilon(\xi,\eta)$ depends only on $\xi$,
$\upsilon(\xi,\eta) = \upsilon(\xi)$.
Equation~(\ref{x8}) reduces then to the Burgers equation.
%which, in Fourier space, takes the form
%
%\begin{equation}
%\label{x12}
%\left( \frac{\partial }{\partial t} + \nu k_\xi^2 \right) \! w(k_\xi) =  -\frac{k_\xi}{2\sqrt{2\pi}}
%\int_{-\infty}^{+\infty} \!\! dq_\xi \, w(q_\xi) \, w(k_\xi-q_\xi).
%\end{equation}
%
The r.m.s. velocity is still given by Eq.~(\ref{x9}), while its spectrum reads
\begin{equation}
\label{x13}
w_{rms}^2(k,t) = 2 \pi k |w(k_\xi,t)|^2,
\end{equation}
where $k = |k_\xi|$. For constant density (case that is not realized in compressible,
pressurless 2HD turbulence), we have $E(t) = v_{rms}^2/2$ (with density equal to one),
so that the energy spectrum is $E(k,t) = w_{rms}^2(k,t)/2 = \pi k |w(k_\xi,t)|^2$.
This result justifies the definition of energy spectrum introduced above.

\section{V. Results}

As discussed in Section II, the state of a turbulent isotropic fluid
depends entirely on the initial conditions, $L_i$ and $u_i$, and on the
dissipation parameter $\nu$. The energy contained in a given wavenumber
$k$ evolves in time with different characteristics in the three
different ranges defined by $k \ll k_i$, $k_i \ll k \ll k_{diss}$,
and $k \gg k_{diss}$. Indeed, while in the large-scale range the energy spectrum
generally depends on the initial conditions but not on dissipation,
in the dissipative range it depends on $\nu$ but not on $L_i$ and $u_i$.
The inertial range, that in wavenumber space is placed in between the two above
ranges, is instead characterized by a {\it universal} energy spectrum
not dependent on either initial conditions or viscosity.
Based on this peculiarities of the energy spectrum, it is plausible
to assume that the wavenumbers defined by $k \ll k_i$ and $k \gg k_i$
``do not communicate'', in the sense that
nonlocal interactions between modes in this different ranges are negligible.
We will assume, in the following, that this is the case.
Accordingly, and roughly speaking, the integral in Eq.~(\ref{cNS})
can be split in two parts
%
%\begin{eqnarray}
%\label{nonlinear}
%\int_{-\infty}^{+\infty} \!\! d\qq \, u(\qq) \, u(\kk-\qq)
%\!\!& = &\!\!\theta(k_i - k) \! \int_{q < k_i} \!\! d\qq \, u(\qq) \, u(\kk-\qq) \nonumber  \\
%\!\!& + &\!\! \theta(k - k_i) \! \int_{q > k_i} \!\! d\qq \, u(\qq) \, u(\kk-\qq), \nonumber  \\
%\end{eqnarray}
%
\begin{equation}
\label{nonlinear}
\int_{-\infty}^{+\infty} \!\! d\qq \, u(\qq) \, u(\kk-\qq)
= \theta(k_i - k) \! \int_{q < k_i} \!\! d\qq \, u(\qq) \, u(\kk-\qq)
+ \theta(k - k_i) \! \int_{q > k_i} \!\! d\qq \, u(\qq) \, u(\kk-\qq),
\end{equation}
where $\theta(x)$ is the Heaviside step function.

{\bf Enstrophy inertial range.} -- In this range,
the energy spectrum has the form~(\ref{Ole5}).
Accordingly, we look for solutions of Eq.~(\ref{cNS}) of the form
\begin{equation}
\label{scaling}
u(\kk,t) = \nu \phi(\kk \sqrt{\nu t}).
\end{equation}
Inserting into Eq.~(\ref{cNS}) and taking into account Eq.~(\ref{nonlinear}),
we find that a possible solution is defined through the scaling
function $\phi(x)$ that, in turns, satisfies the equation
%
%\begin{equation}
%\label{phiEq}
%x \frac{d\phi(x)}{dx} + 2 x^2 \phi(x)  = -
%\frac{i}{\sqrt{2\pi}} \, x \int_{-\infty}^{+\infty} \!\! dy \phi(y) \phi(x-y),
%\end{equation}
%
\begin{equation}
\label{phiEq}
x \frac{d\phi(x)}{dx} + 2 x^2 \phi(x)  = -
\frac{i}{\sqrt{2\pi}} \, x \int_{|y|> k_i/k_{diss}} \!\! dy \phi(y) \phi(x-y),
\end{equation}
where $x = \kk \sqrt{\nu t} = \kk/k_{diss}$. In the inertial range, the
viscous term in not effective and can be neglected. Also, and for the sake
of simplicity, we will take the limit $k_i/k_{diss} \! \rightarrow \!0$
in the bound of the integral in (\ref{phiEq}).
In this case, the Fourier transform of Eq.~(\ref{phiEq}) gives
\begin{equation}
\label{Fourier}
2 \hat{\phi} \hat{\phi}' + \omega \hat{\phi}' + \hat{\phi} = 0,
\end{equation}
where $\hat{\phi}(\omega)$ is the Fourier transform of $\phi(x)$ and a prime
indicates the derivative with respect to $\omega$. Observe that $\hat{\phi}(\omega)$
is exactly the scaling function introduced in Section IV
[see Eq.~(\ref{Burgers3})] and Eq.~(\ref{Fourier}) corresponds to
Eq.~(\ref{Burgers4}) when the dissipative term (proportional to $\hat{\phi}''$)
is neglected.

Equation.~(\ref{Fourier}) admits two solutions
\begin{equation}
\label{Fourier2}
\hat{\phi}(\omega) = -\frac12 \left( \omega \pm \sqrt{\omega^2 + C^2} \, \right)
\end{equation}
specified by the $\pm$ sign, where $C \in \mathbb{R}$ is a constant of integration.
We can neglect the first term in Eq.~(\ref{Fourier2})
since its inverse Fourier transforms is proportional to the derivative
of the $\delta(x)$ function and we are assuming that
$|x| = k/k_{diss} > k_i/k_{diss} \neq 0$. In this case, taking the inverse
Fourier transform of Eq.~(\ref{Fourier2}) gives
\begin{equation}
\label{Fourier3}
\phi(x) = \pm \frac{C}{\sqrt{2\pi} |x|} \, K_1(C|x|),
\end{equation}
where $K_\nu(x)$ is the modified Bessel function of the second kind~\cite{Abramowitz}.

For small values of $|x|$, $C|x| \ll 1$, the asymptotic expansion of $\phi(x)$ is
\begin{equation}
\label{asy}
\phi(x) = \pm \frac{1}{\sqrt{2\pi} x^2}
\end{equation}
at the leading order in $x$. This gives the Batchelor spectrum~(\ref{11})
%
%\begin{equation}
%\label{Eir}
%E(k,t) = c t^{-2} \, k^{-3},
%\end{equation}
%
with $c = 1/2$.

{\bf Dissipative range.} -- This range is defined by the condition $k \gtrsim k_{diss}$.
We can distinguish two ``sub-dissipative ranges'':
In the ``pre-viscous damping range'', the rate of energy transfer due to
nonlinear interactions is not negligible. In this case, we expect a
``slower'' decay of the energy
than the ``fast'' decay in the purely viscous case.
In the ``viscous damping range'', mathematically defined by $k/k_{diss} \rightarrow \infty$,
instead, the rate of energy transfer due to
nonlinear interactions is absent and the dynamics is dominated by viscosity.

{\it Pre-viscous damping range.} -- At the leading order, the asymptotic
expansion of $\phi(x)$ for large $|x|$, $C|x| \gg 1$, is
\begin{equation}
\label{Fourier4}
\phi(x) = \pm \frac{\sqrt{C}}{2 |x|^{3/2}} \, e^{-C|x|} .
\end{equation}
This gives the energy spectrum
\begin{equation}
\label{psiexp}
E(k,t) = \frac{\pi C \nu^{1/2}}{4 k^2 t^{3/2}} \, e^{-2 C k\sqrt{\nu t}}.
\end{equation}
The above result is compatible with the results of Tatsumi and Yanase
discussed in Section II.

{\it Viscous damping range.} -- In this range, we can neglect the
right-hand side of Eq.~(\ref{phiEq}). Accordingly, we get
$\phi(x) = \sqrt{c_\infty/\pi} \, e^{-x^2}$, where $c_\infty > 0$ is a constant.
This gives the energy spectrum~(\ref{2}) in the purely viscous case.

The viscous term is negligible when the second term in Eq.~(\ref{phiEq}) is negligible with
respect to the first one, $\left | d \ln \phi/dx^2 \right| \gg 1$.
%
%\begin{equation}
%\label{vale}
%\left | \frac{d \ln \phi}{dx^2} \right| \gg 1.
%\end{equation}
%
Inserting Eq.~(\ref{Fourier3}), this is the case when
%
%\begin{equation}
%\label{vale2}
%2C|x| \, \frac{K_1(C|x|)}{K_2(C|x|)} \ll C^2.
%\end{equation}
%
$K_1(C|x|)/K_2(C|x|) \ll C/2|x|$ or, using
the asymptotic expansion of $K_\nu(x)$ for large and small
arguments, when
$|x| \lesssim 1$ if $C|x| \lesssim 1$ or $|x| \lesssim C/2$ if $C|x| \gtrsim 1$.
This in turns gives two possibilities for the scaling function
$\psi(|x|) = \pi |\phi(x)|^2$:
\begin{eqnarray}
\label{vale3}
\psi(|x|) =
\left\{ \begin{array}{lll}
\frac{1}{2 x^4} \, , & ~ |x| \lesssim \mbox{min} (1/C,1), \\
c_\infty e^{-2x^2},  & ~ |x| \gtrsim  \mbox{min} (1/C,1),
\end{array}
\right.
\end{eqnarray}
if $C \lesssim \sqrt{2}$ or
\begin{equation}
\label{vale4}
\psi(|x|) =
\left\{ \begin{array}{lll}
\frac{1}{2 x^4} \, ,                      & ~ |x| \lesssim 1/C, \\
\frac{\pi C}{4 |x|^{3}} \, e^{-2C|x|},    & ~ 1/C \lesssim |x| \lesssim C/2, \\
c_\infty e^{-2x^2},                       & ~ |x| \gtrsim C/2,
 \end{array}
\right.
\end{equation}
if $C \gtrsim \sqrt{2}$.
We can get an estimate of $c_\infty$ if we impose the continuity of $\psi(|x|)$ at $|x| = \mbox{min} (1/C,1)$ for the first case ($C \lesssim \sqrt{2}$) and at $|x| = C/2$ for the second case
($C \gtrsim  \sqrt{2}$). Roughly speaking, we get
that $c_\infty$ is of order unity if $C \ll \sqrt{2}$,
and exponentially suppressed, $c_\infty \sim e^{-C^2/2}$, if $C \gg \sqrt{2}$.
%$c_\infty \sim 1$ if $C \ll \sqrt{2}$, and
%$c_\infty \sim e^{-C^2/2}/C^2$ if $C \gg \sqrt{2}$.
%%$c_\infty \sim 1/2e^2$ if $C \ll \sqrt{2}$, and
%%$c_\infty \sim \frac{\pi e^{-C^2/2}}{C^2/2}$ if $C \gg \sqrt{2}$.
%
%\begin{equation}
%\label{cinfty}
%c_\infty \simeq
%\left\{ \begin{array}{lll}
%    \frac{1}{2e^2} \, ,               & ~ C \lesssim 1,     \\
%    \frac{C^4 e^{2/C^2}}{2} \, ,      & ~ 1 \lesssim C \lesssim \sqrt{2}, \\
%    \frac{2\pi e^{-C^2/2}}{C^2} \, ,  & ~ C \gtrsim \sqrt{2}.
%    \end{array}
%\right.
%\end{equation}
%
Accordingly, the exponential tail $c_\infty e^{-2x^2}$ in the spectrum~(\ref{vale4})
is exponentially small when $C$ is large, and then difficult to see both in
experiments and direct numerical simulations of two-dimensional turbulence.

{\bf Large scales.} -- In the limit $k \rightarrow 0$, dissipation is
not effective. Taking into account the Olesen's arguments and
Eq.~(\ref{nonlinear}), we can search for solutions of Eq.~(\ref{cNS})
of the form
\begin{equation}
\label{LS2}
u(\kk,t) = u_i L_i \tau^p \Phi(\ka \tau^q),
\end{equation}
where $\ka = \kk/k_i$, $q = (1+p)/2$, and $\Phi(x)$ is a dimensionless
function of its argument. Inserting the above equation in
Eq.~(\ref{cNS}) and neglecting the term proportional to $\nu$, we find
\begin{equation}
\label{LS6}
q z \, \frac{d\Phi(z)}{dz} + p \Phi(z) = -\frac{i}{2\sqrt{2\pi}} \,
z \int_{-\tau^q}^{+\tau^q} \!\! dz' \Phi(z') \Phi(z - z'),
\end{equation}
where $z = \ka \tau^q$. In the following, and for the sake of simplicity,
we will take the limit $\tau = t/t_i \rightarrow \infty$ in the
bounds of the integral in Eq.~(\ref{LS6}) and we will assume that $q>0$, namely
\begin{equation}
\label{pgreater}
p > -1.
\end{equation}
For $z \rightarrow 0$ (corresponding to $k \rightarrow 0$),
we can write $\Phi(z-z') \simeq \Phi(-z')$. The last assumption
is true if $|\Phi(z)|^2$ is integrable at $z=0$.
Accordingly, to the leading order in $z \rightarrow 0$,
the right-hand side of Eq.~(\ref{LS6}) can be approximated by
$i\Gamma_0 z$, where
$\Gamma_0 = -(1 /2\sqrt{2}\pi )\int_{-\infty}^{+\infty} \!dz |\Phi(z)|^2$
(assuming that the integral exists).
We then find, to the leading order in $z \rightarrow 0$,
\begin{equation}
\label{LS30}
\Phi(z) =
\left\{ \begin{array}{lll}
    3 i\Gamma_0 z \left( C_{-1/3} - i\frac{\pi}{2} + \ln z \right),  & ~ p = -1/3,    \\
    \frac{i\Gamma_0 z}{p+q} + C_p z^{-p/q},                          & ~ p \neq -1/3,
    \end{array}
\right.
\end{equation}
where, $C_{-1/3}$ and $C_p$ and  are dimensionless constants of integration.
The condition $\Phi(-z) = \Phi^*(z)$ implies that $C_{-1/3} \in \mathbb{R}$,
and that $C_p \in \mathbb{R}$ if $p/2q \in \mathbb{Z}$, $C_p$ is purely imaginary
if $(p/q-1)/2 \in \mathbb{Z}$, and
$ \mbox{Im}[C_p]/\mbox{Re}[C_p] = \tan(\pi p/2q) $, otherwise.

For $p = -1/3$, the energy spectrum is given, at the leading order, by
\begin{equation}
\label{LS11}
E(k,t) = c_{-1/3} \, k^3
\ln^2 \!\! \left[ \frac{k}{k_i} \! \left(\frac{t}{t_i}\right)^{\!1/3} \right],
\end{equation}
where $c_{-1/3} = 9\pi u_i^2 L_i^4 \Gamma_0^2$.

For $p \neq -1/3$, we have two cases:

($i$) If $C_p = 0$, the energy spectrum is given by
\begin{equation}
\label{LS18}
E(k,t) = c_p t^{1+3p} k^3,
\end{equation}
where $c_p = \pi u_i^{3(1+p)} L_i^{3(1-p)} \Gamma_0^2/(1+3p)^2$.
The cases $E(k,t) \propto t^\gamma k^3$ with $\gamma = 1, 2, 2.5, 4$ found in the
literature (see~\cite{DavidsonBook}), then correspond to $p =0, 1/3, 1/2, 1$, respectively.
Observe, also, that the energy spectrum at large scales is increasing in time for $p > -1/3$
and decreasing in time for $p < -1/3$.
Finally, observe that for $p = 1$ the energy spectrum does not depend on $L_i$. In this
case $E(k,t) \propto t^4 k^3$. For $p = -1$, instead, the energy spectrum does not depend
on $u_i$. In this case  $E(k,t) \propto t^{-2} k^3$.

($ii$) If $C_p \neq 0$, the integrability condition of $|\Phi(z)|^2$ at $z=0$
implies that $-1 < p < 1/3$. This in turns gives two cases:

($a$) For $-1 < p < -1/3$, we have $-p/q > 1$, so that $\Phi(z) = i\Gamma_0 z/(p+q)$
at the leading order. The energy spectrum is then given by Eq.~(\ref{LS18}).

($b$) For $-1/3 < p < 1/3$, instead, $\Phi(z) = C_p z^{-p/q}$
at the leading order. The energy spectrum is then given by
\begin{equation}
\label{LS13}
E(k,t) = c'_p k^{\alpha_p},
\end{equation}
where $c'_p = \pi u_i^2 L_i^{2(1-p)/(1+p)} |C_p|^2$ and $\alpha_p = (1-3p)/(1+p)$.
Imposing that the Saffman integral is finite we get $\alpha_p \geq 1$,
corresponding to $p \leq 0$. Accordingly, $1 \leq \alpha_p < 3$.

Particularly important is the case $p=0$ (see the Appendix
for a different derivation of the expression of the energy
spectrum in this case). This is the only case where $\Phi(0)$
is finite and different form zero.
In this case, we have the Saffman spectrum
\begin{equation}
\label{LS20}
E(k,t) = c_0 k,
\end{equation}
where $c_0 = \pi u_i^2 L_i^2 |\Phi(0)|^2$.

{\it Analytical case.} -- If one assumes that $\Phi(z)$ is analytical at $z=0$,
namely it admits a Maclaurin expansion, then it is easy to see that the
form of energy spectrum is that given in Eq.~(\ref{expansion})
with $\mathcal{L} = c_0/4\pi$ and
$I(t) = c_p t^{1+3p}$, where $p \neq -1/3$, and
$c_p$ and $c_0$ are the same as those previously defined.

{\bf Absolute state.} -- Let us conclude this section by observing that,
in the inviscid limit, Eq.~(\ref{cNS}) admits a static solution of the form
\begin{equation}
\label{static1}
u(\kk,t) = \frac{e^{i\vartheta\kk}}{\sqrt{a + b k^2}} \, ,
\end{equation}
where $\vartheta \in \mathbb{R}$, $a>0$, and $b>0$ are constants.
This solution corresponds to the static energy spectrum
\begin{equation}
\label{static2}
E(k,t) = \frac{\pi k}{a + b k^2} \, .
\end{equation}
The above spectrum has, correctly, the form of the ``canonical''
distribution for two-dimensional turbulence first discussed by
Kraichnan~\cite{Kraichnan2} (see, also,~\cite{Biskamp}), with $a$
and $b$ related to the inviscid invariants $E$ and $\Omega$.

\section{VI. Conclusions}

In this paper, we have proposed a (continuous) one-dimensional, toy model for
two-dimensional turbulence. Starting by the well-known
(discrete) shell model of hydrodynamics, we have constructed such
a model by adding all possible nonlocal interaction terms
between velocity modes in Fourier space, and then by taking its
continuous limit. In real space, such a model corresponds to
the one-dimensional Burgers equation.

As it is well-known in the literature, Burgers equation
cannot describe the transition from an initial state
to a state of fully developed hydrodynamic turbulence.
This is because the solution of Burgers equation with given Dirichlet
boundary conditions is unable to describe the spontaneous arise of
randomness by chaotic dynamics.

Nevertheless, we have shown that particular asymptotic scaling solutions of Burgers
equation, which do not satisfy Dirichlet
boundary conditions, do indeed well describe the behaviour of a two-dimensional
fluid in a turbulent state, after the transition to chaotic dynamics has happened.
Such scaling solutions successfully explain the main characteristic of the
kinetic energy spectrum at both small, intermediate, and large scales,
as seen in both direct numerical simulations and
laboratory experiments. In particular:

($i$) If the velocity field admits a Maclaurin expansion at small wavenumbers,
then the energy spectrum is that given in Eq.~(\ref{expansion}) as
theoretically predicted, under particular assumptions, in the exact two-dimensional theory.
However, our model also admits the possible existence of a time-independent,
power-law spectrum of the form given in Eq.~(\ref{LS13}).

($ii$) The model predicts the existence of an inertial range,
where the dynamics in independent on initial conditions and viscosity.
Moreover, the energy spectrum is correctly given by the Batchelor spectrum~(\ref{11}).

($iii$) At small scales, corresponding to the dissipative range,
the model admits the existence of two different sub-dissipative ranges.
In the ``pre-viscous damping range'' [see Eq.~(\ref{psiexp})],
where we expect the rate of energy transfer due to nonlinear interactions
to be not negligible, the energy decays slower than in the
``viscous damping range'' [see Eq.~(\ref{2})], where dynamics is dominated
by viscosity.

To our knowledge, the existence of a pre-viscous damping range has been discussed
in the literature only once, in 1981 by Tatsumi and Yanase~\cite{Tatsumi}.
Needless to say, further investigations are needed to confirm the existence and
characteristics of such a peculiar range in freely-decaying two-dimensional
turbulence.

\newpage

\section{Appendix: Saffman spectrum from Burgers equation}

For $p=0$ ($q=1/2$) Eq.~(\ref{LS2}) reads
\begin{equation}
\label{new1}
u(\kk,t) = u_i L_i \Phi(\kk \sqrt{u_i L_i t})
\end{equation}
and, in turns, Eq.~(\ref{LS6}) becomes
\begin{equation}
\label{new2}
z \, \frac{d\Phi(z)}{dz} = -\frac{i}{\sqrt{2\pi}} \,
z \int_{-\infty}^{+\infty} \!\! dz' \Phi(z') \Phi(z - z'),
\end{equation}
where $z = \kk \sqrt{u_i L_i t}$. Taking the Fourier transform of the
above equation, we find
\begin{equation}
\label{new3}
2 \hat{\Phi} \hat{\Phi}' + \varpi \hat{\Phi}' + \hat{\Phi} = 0,
\end{equation}
where $\hat{\Phi}(\varpi)$ is the Fourier transform of $\Phi(z)$ and a prime
indicates the derivative with respect to $\varpi$.
Equation~(\ref{new3}) admits two solutions
\begin{equation}
\label{new4}
\hat{\Phi}(\varpi) = -\frac12 \left( \varpi \pm \sqrt{\varpi^2 + c^2} \, \right)
\end{equation}
specified by the $\pm$ sign, where $c \in \mathbb{R}$ is a constant of integration.
Since the limit $\kk \rightarrow 0$ corresponds to $\varpi \rightarrow \pm \infty$
(large scales), %$\varpi = -\zeta/\sqrt{u_i L_i t}$).
we must take the minus sign in Eq.~(\ref{new4}) in
order to have a finite velocity field and then a finite energy
spectrum. In this case, we have
\begin{equation}
\label{new5}
\hat{\Phi}(\varpi) = \frac{c^2}{4\varpi} + \mathcal{O}(\varpi^{-3}).
\end{equation}
The inverse Fourier transform of the above equation reads
\begin{equation}
\label{new6}
\hat{\Phi}(z) = -\frac{i\sqrt{\pi}c^2}{4\sqrt{2}} \, \mbox{sgn}(z) + \mathcal{O}(z^2),
\end{equation}
where $\mbox{sgn}(x)$ is the sign function.
The leading term of the energy spectrum at large scales is then given by
Eq.~(\ref{LS20}) with $c_0 = \pi^2 c^4 u_i^2 L_i^2/32$. Comparing this
expression for $c_0$ with the one after Eq.~(\ref{LS20}), we find that $c$ is related to
$\Phi(0)$ by $c = 2 (2/\pi)^{1/4} |\Phi(0)|^{1/2}$.

%*******************************   Conflict of Interest   ************************************%

\vspace{1cm}
\begin{center}
{\bf Declarations of interest: none}
\end{center}

%\vspace*{-0.5cm}

%There are no conflict of interest.

%***********************************   Bibliography   ****************************************%

\newpage

%\vspace{0.5cm}
%\begin{center}
%{\bf REFERENCES}
%\end{center} 

\end{document}